\documentclass[aps,prl,reprint]{revtex4-1}
\usepackage{amssymb}
\usepackage{amsmath}
\usepackage{amsthm}
\usepackage{nicefrac}
\usepackage{graphicx}
\usepackage{epstopdf}
\usepackage{xcolor}

\begin{document}

\title{Sudden viscous dissipation of compressing turbulence
}

\author{Seth Davidovits}
\author{Nathaniel J. Fisch}
\affiliation{Princeton University, Princeton Plasma Physics Laboratory, Princeton, New Jersey 08540, USA}

\begin{abstract}
Compression of  turbulent plasma  can  amplify  the turbulent kinetic energy, if the compression is fast compared to the viscous dissipation time of the turbulent eddies. A sudden viscous dissipation mechanism is demonstrated, whereby this amplified turbulent kinetic energy is rapidly converted into thermal energy, suggesting a new paradigm for fast ignition inertial fusion.
\end{abstract}

\maketitle

\emph{Introduction.}---Unprecedented densities and temperatures are now reached by compressing plasma using lasers or magnetic fields, with the objective of reaching nuclear fusion, prodigious x-ray production, or new regimes of materials. 
The plasma motion in these compressions can be turbulent, whether magnetically driven \cite{kroupp2007,kroupp2007a,kroupp2011,maron2013} or laser driven \cite{thomas2012,weber2014}. 
However, rapid compression of this turbulent plasma,  where the viscosity is highly sensitive to temperature, is demonstrated here to exhibit unusual behavior, where the turbulent kinetic energy (TKE) abruptly switches from growing to rapidly dissipating. This behavior occurs in plasma, but is not predicted by studies of neutral gas compression \cite{wu1985,coleman1991,cambon1992,durbin1992,coleman1993,cambon1993,hamlington2014}.
In fact,  it was  observations of the dominant effect of the TKE in Z-pinches, both in pressure balance  \cite{maron2013} and  in radiation balance   \cite{kroupp2007a,maron2013},  that stimulated the present study.

Compression is rapid if the rate of compression is fast compared to the turbulent timescale $\tau = k/\epsilon$ with $k$ the TKE and $\epsilon$ the viscous dissipation rate. 
In the initially rapid plasma compressions here, the viscous dissipation eventually grows such that the turbulent timescale $\tau$ shortens, and the plasma TKE suddenly dissipates. 
This dissipation is sudden in that it occurs over a time interval that is small compared to the overall compression time. 

This sudden dissipation mechanism now suggests a new fast ignition paradigm.
Imagine an initially turbulent fusion fuel plasma where the majority of the energy is in the turbulent motion. 
This plasma is then rapidly compressed, causing both the TKE and thermal energy to grow, while the TKE retains most of the energy, as observed in certain Z-pinch experiments \cite{kroupp2007,kroupp2007a,kroupp2011,maron2013}.
Since  radiation losses (both synchrotron and bremsstrahlung) and nuclear fusion are dependent on thermal energy but not the TKE,  those processes are minimized under such compression, since the plasma stays comparatively cool.
However, late in the compression, the TKE suddenly dissipates viscously into thermal energy, thereby igniting the plasma without having undergone large radiation losses. 

In neutral gas, upon  rapid compression, the TKE grows. 
In an isotropic 3D compression, it grows as $1/L^2$, where $L$ is the (time dependent) side length of a box that is compressing with the mean flow along each axis. 
This is true for both the zero Mach case \cite{wu1985}, where the TKE is solenoidal, as well as in the finite Mach case, where the TKE has both solenoidal and dilational components, which each grow in energy as $1/L^2$ \cite{cambon1993}. 
This is the same rate at which the temperature increases for an adiabatic ideal gas compression in 3D. 
Thus, in an idealized rapid compression the ratio of TKE to thermal energy stays constant, making the initial conditions very important for the energy dynamics while the compression is rapid.

In plasma, as in neutral gases, the TKE similarly grows under compression, but the greater sensitivity of viscosity to temperature now has a telling effect.
The rapid compression causes the TKE to grow, with most of the TKE contained  in the large scale eddies. 
The energy equation for the viscous dissipation of TKE, $E_{\rm TKE}$, is
\begin{equation}
\frac{\mathrm{d} E_{\rm TKE}}{\mathrm{d} t} = \frac{\mu\!\left(T\right)}{\rho} \langle \mathbf{u}_l \cdot \nabla^2 \mathbf{u}_l \rangle \sim - L T^{5/2} E_{\rm TKE}, \label{eq:energy sketch}
\end{equation}
where the last scaling reflects the viscosity dependence in plasma going as $\mu\!\left(T\right) \sim T^{5/2}$. This contrasts with the weaker scalings used for compressing gases, such as $\sim T^{3/4}$ (e.g. \cite{wu1985,coleman1993}).
The compression forces the energy containing eddies to smaller and smaller scales; $\nabla^2 \sim 1/L^2$, where $L$ is the shrinking domain (and largest eddy) scale. 
The density scales as $\rho \sim 1/L^3$. 
The temperature increases during the compression, going as (for a 3D adiabatic compression of monatomic gas) $T \sim 1/L^2$,
so that the total dissipation scales as $1/L^4$. 
The increasing viscosity dissipates the smallest scales first, then works up to the large energy-containing scales.
The sensitive dependency of the viscosity to the temperature now means that the viscous dissipation of the large-scale TKE occurs rapidly as a function of $L$.
Thus, under rapid compression, the TKE first grows with decreasing $L$, and then suddenly dissipates essentially at constant $L$.

This sudden dissipation effect is demonstrated here numerically in the limit of small initial TKE, so that the increase in temperature is due only to the direct compression, and not the self-consistent transfer of TKE to thermal energy.  
For turbulence where a substantial fraction of the energy is contained in the TKE, the dissipation should be even more sudden, in fact, explosive,
because there is a feedback mechanism for the viscous dissipation. 
To see this, consider that  the viscous dissipation drives the temperature according to: $dT/dt \sim LT^{5/2} E_{\rm TKE}$. 
If most of the energy is in the TKE, then the  viscous dissipation  will rapidly raise the temperature. 
This in turn, will rapidly raise the viscosity, since  $\mu\!\left(T\right) \sim T^{5/2}$, accelerating the conversion of TKE into temperature.
The result is an explosive instability, until the TKE is depleted.  
Thus, while we simulate sudden viscous dissipation for small initial TKE, it is with the understanding that the effect may be even more sudden for large  initial TKE.
The sudden dissipation in any event underlies the proposed new paradigm for  fast ignition.

\begin{figure*}
\includegraphics[]{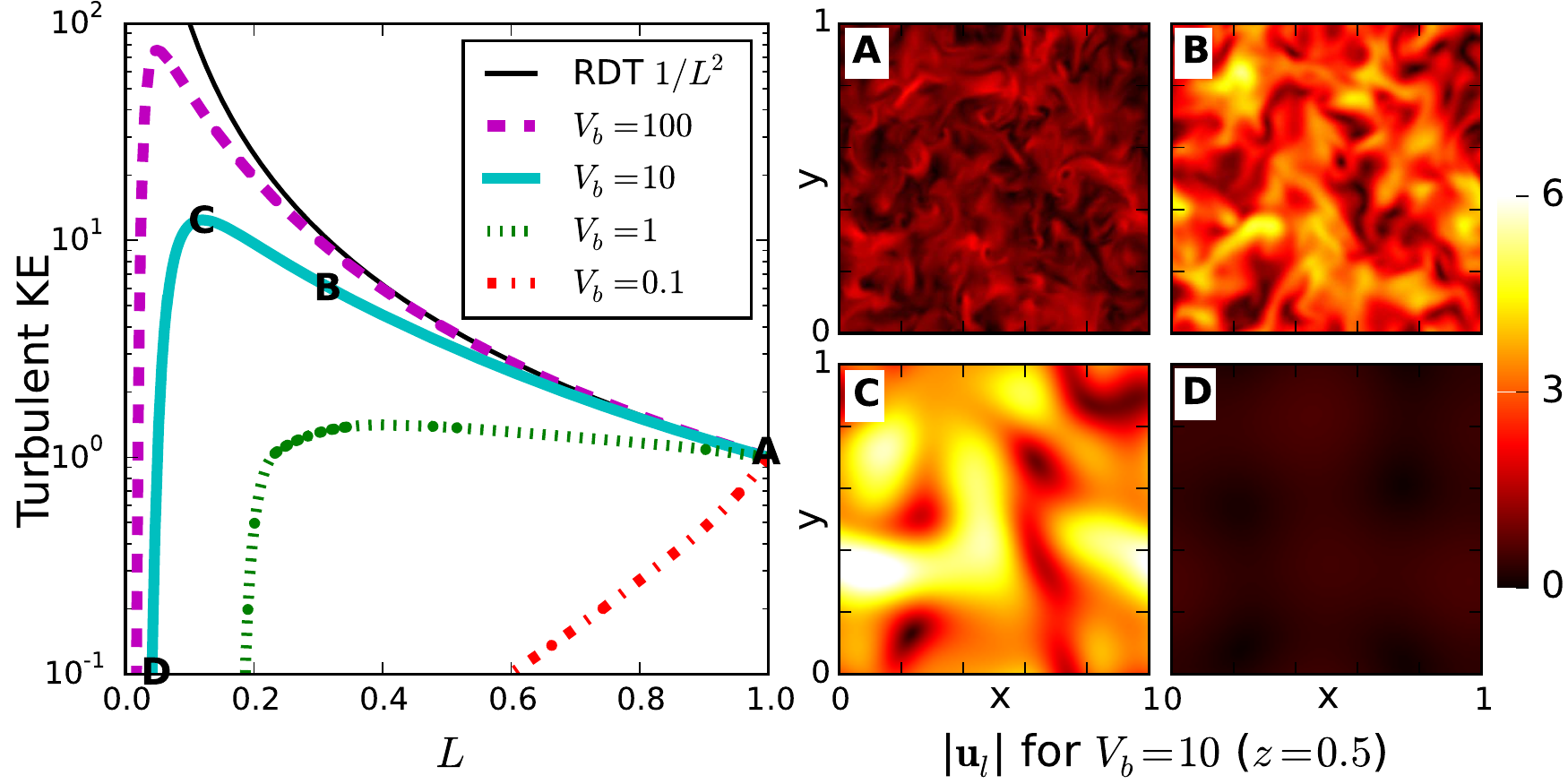}
\caption{LEFT: Turbulent kinetic energy (TKE) during compression at various rates. An initially turbulent flow field, with TKE normalized to 1, is compressed on times ($\tau_c = 1/2V_b$) slower than ($V_b=0.1$), comparable to ($V_b=1$) and faster than ($V_b = 10,100$) the initial turbulence timescale $\tau_0=(k/\epsilon)_0\sim1/2$. The initial domain is a box of size $1^3$, time progresses right to left ($t=(1-L)/(2V_b)$) as the compression shrinks the domain. When the compression is slow, the TKE damps, albeit at a slower rate than it would with no compression. When it is comparable, the energy stays relatively constant before damping. When it is faster, the TKE grows substantially, before being suddenly dissipated over a small range of L. Also shown is the theoretical rapid distortion theory (RDT) solution, which is exact while the compression is extremely fast.
RIGHT: Flowfield slices showing the magnitude of the turbulent flow velocity in the lab frame for $V_b=10$. The fields progress in time from A$\rightarrow$B$\rightarrow$C$\rightarrow$D, and are marked on the left graph to show the amount of compression and the total TKE of the flow. As time increases the flow is increasingly contained in the largest structures as the smaller structures run into the viscous scales; see also Fig. \ref{fig:Ek}. All plots are normalized to side length 1, in the lab frame the absolute size of all structures decreases in time.
\label{fig:KEvsL}}
\end{figure*}

\begin{figure}
\includegraphics[]{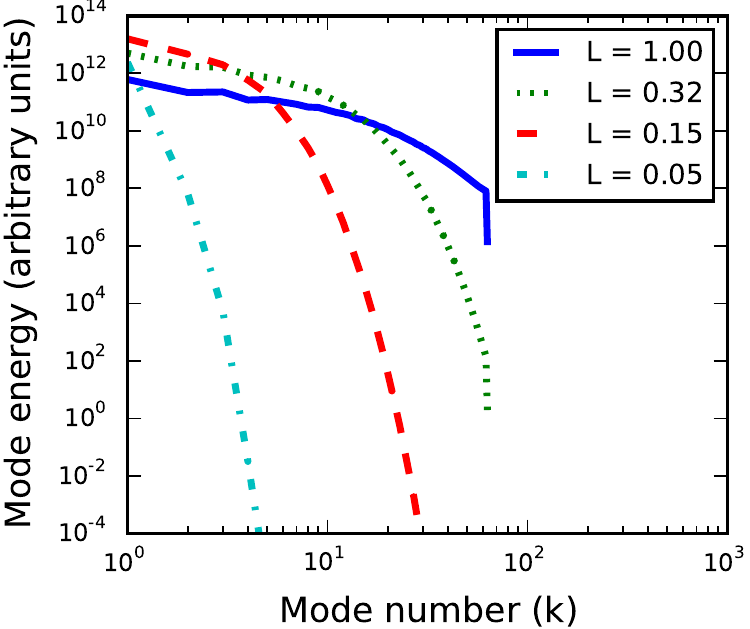}
\caption{Fourier mode distribution of the turbulent kinetic energy (TKE) for turbulence compressed at a rate that is initially fast compared to the turbulent timescales ($V_b = 10$). The total TKE as a function of L (time) is shown as the solid cyan line in Fig.~\ref{fig:KEvsL}. The initial spectrum is $L=1$ (blue, solid). After  moderate  compression ($L=0.32$, dotted green), but before the peak TKE, the energy in the highest modes has damped substantially, but energy has gone into the large scale flow. Near the peak TKE ($L=0.15$, dashed red) this trend has continued, with viscosity working its way towards larger scales, and the largest scales continuing to gain energy. When the viscosity hits the largest scales, the energy is suddenly dissipated, after which all modes have been damped ($L=0.05$, dash-dotted cyan). The four spectra approximately correspond in L (time) to the four fields in Fig. \ref{fig:KEvsL}. \label{fig:Ek}}
\end{figure}

\emph{Model.}---
To model the compression of a turbulent plasma in the zero Mach limit, we follow previous work on compressing fluids \cite{wu1985,blaisdell1991,coleman1991,cambon1992,hamlington2014}. The total (subscript $t$) plasma density, pressure, and flow are decomposed into mean (overbar) and fluctuating parts (subscript $l$),
\begin{subequations}
\label{eq:splittings}
\begin{align}
\rho_t &= \bar{\rho} + \rho_l, \label{eq:rho split} \\
p_t &= \bar{p} + p_l, \label{eq:p split}\\
\mathbf{u}_t &= \bar{\mathbf{u}} + \mathbf{u}_l. \label{eq:u split}
\end{align}
\end{subequations}
The mean flow, $\bar{\mathbf{u}}$ in Eq.~(\ref{eq:u split}) is taken to be an externally enforced background flow of the form
\begin{equation}
\bar{\mathbf{u}} = A\!(t) \mathbf{x}_l = (\dot{L}/L) I \mathbf{x}_l,
\end{equation}
where I is the identity matrix, the subscript $l$ on the coordinate $\mathbf{x}_l$ is to indicate it is a stationary lab coordinate, and $L\!(t)$ is the length along each side of a box that is advected with the background flow. The compression is chosen to be isotropic, so that the matrix A can be rewritten with I, and $\dot{L}$ is taken to be constant.

In the low Mach number limit, the density fluctuations $\rho_l$ are neglected, and the turbulent velocity is divergence free $\mathbf{u}_l$ \cite{blaisdell1991,cambon1992}. The mean density depends only on time,
\begin{equation}
\bar{\rho} = \rho_0 \left(L_0/L\!(t)\right)^3, \label{eq:density evolution}
\end{equation}
with $\rho_0$ and $L_0$ the initial density and box length respectively. Under these conditions, the momentum equation for the plasma has the form \cite{cambon1992},
\begin{equation}
\frac{\partial \mathbf{u}_l}{\partial t} + \mathbf{u}_l \cdot \nabla_l \mathbf{u}_l + A\mathbf{x_l}\cdot \nabla_l \mathbf{u}_l + A\mathbf{u}_l = \frac{-1}{\bar{\rho}} \nabla_l p_l + \frac{\mu\left(T\right)}{\bar{\rho}} \nabla_l^2 \mathbf{u}_l. \label{eq:lab momentum}
\end{equation}
The viscosity is taken as a function of temperature,
\begin{equation}
\mu\!\left(T\!\left(t\right)\right)= \mu_0 \left(T\!\left(t\right)/T_{0}\right)^{\beta}= \mu_0 \left(L_0/L\!\left(t\right)\right)^{2\beta}.
\end{equation}
Here $T_0$ and $\mu_0$ are the initial temperature and the initial (dynamic) viscosity respectively. The compression is assumed adiabatic, and the gas is ideal and monatomic, so that the temperature increases as $T\!\left(t\right) = T_0 (L_0/L\!\left(t\right))^2$. As previously stated, feedback from the viscous dissipation of kinetic energy into temperature is neglected.

Working in a frame that moves with the background flow $\bar{\mathbf{u}}$ eliminates the explicit $\mathbf{x}_l$ dependence in Eq.~(\ref{eq:lab momentum}), allowing for periodic boundary conditions. This is achieved with the coordinate transformation $\mathbf{x}_{l}= L\!(t) \mathbf{x}.$ The turbulent velocity is transformed as
\begin{equation}
\mathbf{u}_l \! (\mathbf{x}_l,t) = \mathbf{u} \! (\mathbf{x},t). \label{eq:velocity transformation}
\end{equation}
and the pressure is transformed similarly, $p_l \! (\mathbf{x}_l,t) = p \! (\mathbf{x},t)$.

After transformation, the momentum equation for $\mathbf{u}$ is
\begin{equation}
\frac{\partial\mathbf{u}}{\partial t}+\frac{1}{L}\mathbf{u}\cdot\nabla\mathbf{u}+\frac{\dot{L}}{L}\mathbf{u}=-L^2\nabla p+\frac{1}{\mbox{Re}_{0}}L\mu(T)\nabla^{2}\mathbf{u} \label{eq:momentum mf}
\end{equation}
where the standard nondimensionalization has been used, and $L$ is understood to be normalized to $L_0$. The Reynolds number is subscripted with a $0$ because one may view the effective Reynolds number as changing due to the compression.

The effects of the compression in the moving frame appear as time dependent coefficients on the nonlinear, pressure and dissipation terms, and as a new term, proportional to $\mathbf{u}$. This new term may be viewed as a forcing or damping, depending on its sign (as written, a negative sign is forcing). Indeed, a (constant coefficient) term proportional to $\mathbf{u}$ has been used to force turbulence for turbulence codes working in real space, where spectral forcing methods may be difficult to implement \cite{lundgren2003,rosales2005,carroll2013}. 

The variables in the momentum equation in the moving frame, Eq.~(\ref{eq:momentum mf}) can be rescaled so that the forcing term is eliminated, and one is left with regular Navier-Stokes with a time dependent viscosity \cite{cambon1992}. We choose instead to eliminate the time dependence from all terms except the forcing term, by using the scalings,
\begin{subequations}
\label{eq:scalings}
\begin{align}
\mathbf{u} &=  L^{\left(2 - 2 \beta \right)}\hat{\mathbf{u}}, \label{eq:u scaling} \\
p &= L^{\left(1 - 4 \beta \right)}\hat{p}, \label{eq:p scaling} \\
{d}t' &= L^{\left(1 - 2 \beta \right)}\mbox{d}t. \label{eq:t scaling}
\end{align}
\end{subequations}
Doing so speeds up our simulations, by eliminating time dependence from the stiff viscous term. The momentum equation, Eq.~(\ref{eq:momentum mf}), with the scalings in Eqs.~(\ref{eq:scalings}) is,
\begin{equation}
\frac{\partial \hat{\mathbf{u}}}{\partial t'}+\hat{\mathbf{u}}\cdot\nabla\hat{\mathbf{u}}+(3-2\beta)\frac{\dot{L}}{L^{2-2\beta}} \hat{\mathbf{u}}=-\nabla\hat{p}+\frac{1}{\mbox{Re}_{0}}\nabla^{2}\hat{\mathbf{u}}. \label{eq:final momentum}
\end{equation}
The scaled momentum equation in the moving frame, Eq.~(\ref{eq:final momentum}), is incompressible Navier-Stokes with an extra time dependent forcing (or damping) term proportional to $\hat{\mathbf{u}}$. 

\emph{Sudden viscous dissipation.}---
We consider constant velocity compressions, with $L = 1 - 2 V_b t$, and use the plasma viscosity, $\beta=5/2$. The forcing term in Eq.~(\ref{eq:final momentum}) acts as a damping term, the strength of which decays in time, asymptoting to 0 as $t' \rightarrow \infty$ ($L \rightarrow 0$). 
Thus, in the moving frame and scaled variables, the turbulence simply decays, at a rate that is faster than through viscosity alone. The damping term, in Fourier (wavenumber) space, has no wavenumber dependence and damps all modes proportional to their strength, so that the overall spectral dynamics is different than decaying turbulence. 

While in this frame and variables the turbulence decays, in the lab frame, it may grow. Equation (\ref{eq:u scaling}) combined with Eq.~(\ref{eq:velocity transformation}) show the lab flow is amplified by the factor $1/L^3$ compared to the moving frame flow. 
Thus, even though the flow field $\hat{\mathbf{u}}$ decays in the moving frame, it may increase in the lab frame if the decrease in $L$ is rapid enough that the $1/L^3$ growth overcomes the decay. 

We simulate Eq.~(\ref{eq:final momentum}) with periodic boundary conditions using the spectral code Dedalus \cite{dedalus}. 
An initial turbulent flow field is generated using the forcing method of Lundgren \cite{lundgren2003,rosales2005}, with a coefficient of 1 so that the initial turnover time is approximately $1/2$, and $\mbox{Re}_0$ is taken as 3000. 
We use a $128^3$ Fourier grid and 3/2 dealiasing. 
The simulations are initially under-resolved, but quickly become resolved as the compression progresses. 
Our main focus is to show the qualitative behavior of the sudden dissipation. 
For the same initial flow field, we simulate compression at various compression velocities $V_b$. The compression is continued until the turbulent kinetic energy dissipates. Different initial flow fields with similar but non-identical energy spectra show similar behavior to the results presented. Key results are shown in Figs.~\ref{fig:KEvsL} and \ref{fig:Ek} and discussed in the captions. 

To understand this sudden dissipation,  consider the TKE equation in the lab frame. Because $\mathbf{u}_l = \mathbf{u}$, Eq.~(\ref{eq:velocity transformation}), the lab TKE can be put as $k\!\left(t\right) = \langle \mathbf{u}^{2}/2 \rangle$,
where, for the homogeneous, isotropic case considered here, the angle brackets denote a spatial average over the domain. The equation for $\mathrm{d}k/\mathrm{d}t$ is
\begin{equation}
\frac{\mathrm{d}k}{\mathrm{d}t}=-2\frac{\dot{L}}{L}k - L^{1-2\beta}\epsilon. \label{eq:KE eq}
\end{equation}
Here, $\epsilon$ is the viscous dissipation, given by $\epsilon  =  -\langle\mathbf{u}\cdot\nabla^{2}\mathbf{u}\rangle/\mathrm{Re}_{0}$,
and the first term on the right hand side in Eq~(\ref{eq:KE eq}) represents energy increase due to the compressive forcing. The compression is rapid when this energy increase term is much larger than the viscous dissipation term. Then the dissipation term in Eq.~(\ref{eq:KE eq}) can be dropped and the kinetic energy in the lab frame is found to be $k\!\left(L\right) = k_0/L^2$. This is the basic rapid distortion theory solution \cite{wu1985} which is shown for comparison in Fig.~\ref{fig:KEvsL}. When $\beta=5/2$, the dissipation term in Eq.~(\ref{eq:KE eq}) has a prefactor of $1/L^4$, which grows strongly as L decreases. It is apparent from Fig.~\ref{fig:KEvsL} that under rapid compression the turbulent kinetic energy initially grows, but at some point the growth of the viscous dissipation overcomes this growth and causes it to damp. 

\emph{Discussion and Caveats.}---
For an illustration of a parameter regime where the sudden dissipation effect could possibly appear, consider the parameters of a magnetized liner inertial fusion (MagLIF) point design \cite{slutz2010}; after laser pre-heating, but before compression, the cylindrical liner with radius 2.7 mm contains a 50/50 mix of deuterium (D) and tritium (T) at 250 eV, and densities of $\sim 3.5\times10^{20}/\rm{cm}^3$. The gas has a kinematic viscosity of $\nu \sim 7.8\times10^{-3}\,\rm{m}^2/\rm{s}$. Suppose the gas had an initial Reynolds number of 3000, as used for the results in Fig.~\ref{fig:KEvsL}. This would mean large scale flows on order of $8.6\,\rm{km}/\rm{s}$ existed inside the capsule, which would be much smaller than the D and T thermal velocities ($\sim 100 \rm{km}/\rm{s}$), making the turbulence low Mach.  The Z machine compresses the gas with a velocity around 50 km/s, slower than the thermal speeds, but much faster than the flow speed. Approximating it as 10 times faster, making crude accounting for the fact the compression is 2D rather than 3D, and comparing to the cyan curve on Fig.~\ref{fig:KEvsL}, it is estimated that the flow energy would peak and begin suddenly dissipating after a factor of 10 compression, around a radius of 0.27 mm. The TKE would grow up by a factor of approximately 10 before the dissipation, while remaining a minor fraction of the total energy. A more complete plasma model should be used before predictions are made for any specific experiment, MagLIF included.

There are, of course, a number of important caveats for the present work. 
On one hand, very precise measurements show that compression on the WIS Z-pinch leads to ion motion that is dominated by TKE in the imploding plasma \cite{kroupp2007}. 
On the other hand, alternative explanations  for this effect have not been exhausted \cite{giuliani2014}. 
Also, the proposed fast ignition paradigm contemplates supersonic turbulence, not the zero Mach number turbulence limit simulated here. 
A flow with much more energy in the turbulence than in temperature is necessarily supersonic. 
While the present work provides good motivation for plasma with supersonic TKE also to exhibit a sudden viscous dissipation effect, it remains to demonstrate it. As noted in the introduction, in the small Mach number turbulence limit, rapid 3D compression causes the TKE to grow as $1/L^2$, as does the temperature. Therefore the turbulence in such compressions stays low Mach. The initial state of supersonic turbulence, which decays on a timescale long enough to be rapidly compressed, must be generated in some fashion.  A staged implosion, where a first stage generates supersonic turbulence, and a second stage rapidly compresses it, is one possibility.

The low Mach number in the compressions treated here means that even at late times, when the Reynolds number is small, the Knudsen number is small ($\rm{Kn} \sim \rm{M}/\rm{Re}$), so that the fluid description should remain appropriate.

Note that the constant velocity compression considered here requires an ever increasing compressive force. 
This may occur naturally for some period of time (not indefinitely), especially on Z-pinches where the constriction of the current causes the magnetic pressure to continually increase. 
The consideration of alternate compression histories is straightforward. Boundary effects, such as viscous drag on the fluid motion, were ignored here through the use of periodic boundaries. They may be important in practical applications, especially as the imploding fuel becomes smaller. In the late stages of the implosion a new equation of state and viscosity dependence on temperature may alter the results.
Finally, although the virtuous features of TKE were exploited here, namely the reduced radiation and the deferral of fusion reactions,  there could be deleterious features of most of the energy residing in TKE during the compression phase, such as by increasing heat transport  or contributing to mixing \cite{wilson2003,thomas2012,weber2014}.  

These caveats notwithstanding, what remains is a tantalizing, if speculative, new paradigm, namely  using rapid compression to increase the energy in turbulent structures incapable of radiating much energy away during the compression, nor prematurely igniting the plasma through fusion reactions, with the energy in these turbulent structures then explosively converting to heat as the viscosity grows, thereby creating the deferred ignition conditions without radiative loss. 
The strong dependency of viscosity on temperature in plasma facilitates the sudden dissipation, as confirmed by the simulations presented here in the subsonic limit. 

\begin{acknowledgments}
This work was supported by DOE through contracts DE-AC02-09CH1-1466 and 67350-9960 (Prime $\#$ DOE DE-NA0001836) 
and by DTRA HDTRA1-11-1-0037.  For stimulating the ideas presented here, the authors would like to acknowledge the discovery of the dominant effect of TKE in compressing Z-pinch experiments carried out at the Weizmann Institute   \cite{kroupp2007,kroupp2007a,kroupp2011,maron2013}.
\end{acknowledgments}

\vfill\eject

\end{document}